# Synthesis, structural and magnetic properties of the solid solution (CuCl$_{1-x}$Br$_x$)LaNb$_2$O$_7$ (0 ≤ x ≤ 1)

Yoshihiro TSUJIMOTO*, Atsushi KITADA, Hiroshi KAGEYAMA**, Masakazu NISHI[1], Yasuo NARUMI[1, 2], Koichi KINDO[1], Yoko KIUCHI[1], Yutaka UEDA[1], Yasutomo J. UEMURA[3], Yoshitami AJIRO and Kazuyoshi YOSHIMURA

Department of Chemistry, Graduate School of Science, Kyoto University, Kyoto 606-8502

[1] Institute for Solid State Physics, University of Tokyo, 5-1-5 Kashiwanoha, Chiba 277-8581

[2] Institute for Materials Research, Tohoku University, Katahira 2-1-1 Aoba-ku, Sendai 980-8577

[3] Department of Physics, Columbia University, New York 10027, USA

*e-mail: yoshi@kuchem.kyoto-u.ac.jp

** e-mail: kage@kuchem.kyoto-u.ac.jp

**Abstract**

The two-dimensional (2D) quantum spin system (CuCl)LaNb$_2$O$_7$ has a spin-singlet ground state with a gap of 2.3 meV, while the isostructural material (CuBr)LaNb$_2$O$_7$ displays collinear antiferromagnetic order at $T_N$ = 32 K. Here we report on the synthesis of solid solution (CuCl$_{1-x}$Br$_x$)LaNb$_2$O$_7$ (0 ≤ $x$ ≤ 1), and its structural and magnetic properties by means of magnetic susceptibility, high-field magnetization, and neutron diffraction studies. The $x$-dependence of cell parameters follows Vegard's law, verifying the uniform distribution of Cl and Br atoms at the halide site, though a more complex structural evolution is inferred from an opposing correlation between the intra- and interlayer cell distances (vs. $x$). 5%-Br substitution is found to induce antiferromagnetic order with $T_N$ = 7 K, consistent with a recent $\mu$SR study, and the magnetic structure is collinear, having a significantly reduced moment. Further Br substitution leads to a linear increase in $T_N$ up to $x$ = 1. These results indicate that (CuCl)LaNb$_2$O$_7$ is located in the vicinity of the quantum phase boundary.

Novel quantum phase transitions can be driven by tuning various parameters such as magnetic field and pressure, instead of temperature. For a quantum spin system having a spin-singlet ground state with a finite gap, application of an external magnetic field is known to induce Bose-Einstein condensation (BEC) of triplet magnons as observed in $TlCuCl_3$ [1−4], or crystallization of triplet magnons to yield magnetization plateaus as observed in $SrCu_2(BO_3)_2$ [5].

Cationic substitution is also a useful and versatile parameter for spin-singlet compounds, which is classified into two types. One is substituting the magnetic site by (non)magnetic ions with different spin multiplicity. Magnetic order is induced by a small amount of $Zn^{2+}$ ($S = 0$) substitution for $Cu^{2+}$ ($S = 1/2$) in $CuGeO_3$ [6] and $SrCu_2O_3$ [7], and $Mg^{2+}$ ($S = 0$) substitution for $Ni^{2+}$ ($S = 1$) in $PbNi_2V_2O_8$ [8]. Of particular interest is microscopic phase segregation in a doped one-dimensional chain between spin-ordered and disordered states with a modulated staggered magnetization along the chain [9, 10]. The other type of cation substitution is conducted for the counter cation, which provides chemical pressure and/or bond randomness in the magnetic coupling constants. A Bose-glass state has recently been suggested in $Tl_{1-x}K_xCuCl_3$ [11, 12]. By contrast, anionic substitution (O-S-Se, F-Cl-Br) has been rarely studied, mainly due to a large difference in anionic radius and in electronegativity, making it difficult to prepare a homogeneous solid solution. However, when magnetic ions are bridged by substitutable anions, we can control the superexchange coupling constants.

$(CuX)LaNb_2O_7$ ($X$ = Cl, Br), as prepared by ion-exchange reactions, consist of $CuX$ square-lattice layers with $S = 1/2$ separated by non-magnetic perovskite blocks (see Fig. 1) [13]. $(CuCl)LaNb_2O_7$ has a spin-singlet ground state separated from the excited sate by 2.3 meV [14], and magnetic field induces the Bose Einstein condensation (BEC) of triplet magnons at 10.3 T [15, 16]. In contrast, the Br counterpart undergoes collinear antiferromagnetic (CAF) order at 32 K [17]. These compounds were initially assumed as candidates of the $J_1$-$J_2$ model where the ferromagnetic (FM) nearest-neighbor ($J_1$) and the antiferromagnetic (AFM) next-nearest-neighbor ($J_2$) interactions compete with respect to each other [15, 18]. However, nuclear magnetic resonance (NMR) and transmission electron microscopy studies on he two compounds demonstrated the absence of the $C_4$ symmetry at both Cu and $X$ sites [19], inconsistent with the original structural analyses [13]. Several superstructures in the $CuX$ plane were proposed to explain the spin-singlet state for $X$ = Cl [19] and the CAF state for $X$ = Br [20]. But it is not yet clear whether the proposed model can answer the questions, for example, of why the field-induced transition occurs at much lower field than that expected from the zero-field spin gap, and how magnetic coupling differs between $X$ = Cl and Br.

In this study, we present successful preparation of the whole solid solution $(CuCl_{1-x}Br_x)LaNb_2O_7$ ($0 \leq x \leq 1$) and structural and magnetic characterizations by means of X-ray diffraction, neutron diffraction, magnetic susceptibility and high-field magnetizations. The motivation of the present study is partly due to the recent observation of magnetic order by muon spin relaxation ($\mu$SR) for a 5% Br-substituted sample ($x$ = 0.05) [21].

Preparation of $(CuCl_{1-x}Br_x)LaNb_2O_7$ ($x$ = 0, 0.05, 0.33, 0.50, 0.66, 1) was performed by the following ion-exchange reaction:

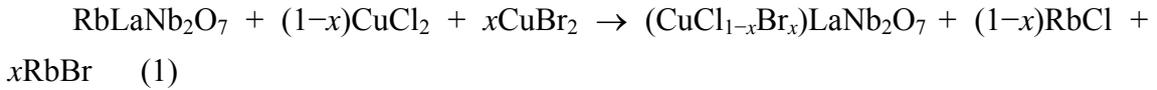

$RbLaNb_2O_7$ + (1−$x$)$CuCl_2$ + $x$$CuBr_2$ → $(CuCl_{1-x}Br_x)LaNb_2O_7$ + (1−$x$)RbCl + $x$RbBr     (1)

The parent compound $RbLaNb_2O_7$ was prepared by a high-temperature reaction using $Rb_2CO_3$ (Rare Metallic, 99.9 %), $La_2O_3$ (Rare Metallic, 99.99%), and $Nb_2O_5$ (Rare Metallic, 99.99%) following the literature [22]. Anhydrous $CuCl_2$ (Alfa, 99.999) and $CuBr_2$ (Alfa, 99.999/99.9 %) in a stoichiometric ratio, $x$, were mixed with $RbLaNb_2O_7$ pressed into pellets inside an argon-filled glove box, sealed in evacuated (<$10^{-3}$ Torr) Pyrex tubes and heated at 320 °C for 7 days. To complete the desired reaction, a two-molar excess of $CuCl_2$/$CuBr_2$ was used relative to $RbLaNb_2O_7$. The final products were isolated by washing with warm distilled water.

X-ray powder diffraction measurements were carried out at room temperature with Cu $K_\alpha$ radiation on a M18XHF diffractometer (Mac Science). Magnetic susceptibility measurements were performed using a superconducting quantum interference device (SQUID) magnetometer (Quantum Design, MPMS) over a temperature range 2 − 300 K in an applied field $H$ = 0.1 T. High-field magnetization measurements were conducted using an induction method with a multilayer pulsed magnet installed at the Institute for Solid State Physics, the University of Tokyo (ISSP). Magnetization data were collected at 1.3 K in magnetic fields up to 60 T. Powder neutron diffraction was performed on the 5% Br-substituted sample using the ISSP-PONTA triple-axis spectrometer (5G) installed at JRR-3 of the Japan Atomic Energy Agency (JAEA), Tokai. A polycrystalline sample of about 10 g mass was placed in a vanadium cylinder. Neutrons with a wavelength of 2.3618 Å were obtained by the (002) reflection of pyrolytic graphite (PG), and a collimation of open-40'-sample-80'-80' was employed in combination with a PG filter placed before the sample to eliminate higher-order beam contamination.

Elemental analysis of the products, carried out by energy dispersive spectroscopy (EDS) on a JSM-5600 scanning electron microscope (JEOL), supported the nominal stoichiometry, $(CuCl_{1-x}Br_x)LaNb_2O_7$. X-ray diffraction profiles of the whole solid solution $(CuCl_{1-x}Br_x)LaNb_2O_7$ were indexed to the tetragonal symmetry as in the case of the end members ($x = 0, 1$) [13]. The peak widths of the solid solution were as sharp as those of the end members, and no multiphase region was observed within the experimental resolution. Uniform distribution of the Cl and Br atoms to the $X$ site was further supported by EDS on a JEM-2010F transmission electron microscopy (TEM) with an operating voltage of 200 kV at ISSP. The $x$ dependence of the lattice parameters as shown in Fig. 2 demonstrates that both $a$ and $c$ parameters change in proportion to $x$, following Vegard's law, implying successful preparations of the whole solid solution. However, the Br substitution cannot be simply interpreted as due to the negative chemical pressure, since the $a$ axis increases and the $c$ axis decreases with increasing $x$, respectively, a finding whose origin is not clear at the present stage. But this should be a common feature in this family of ion-exchanged compounds, as $(CuBr)LaTa_2O_7$ and $(CuBr)La_2(Ti_2Nb)O_{10}$ also have shorter $c$ axes than their Cl-counterparts [23]. Determination of the precise structure is required in the future. For example, a superstructure, as found in $(CuCl)LaNb_2O_7$ [19], might also be observed in the (Cl, Br) solid solution.

Figure 3(a) presents the temperature dependence of the magnetic susceptibility $\chi(T)$ for the solid solution in the region below 50 K. The kink at 32 K for $x = 1$ was ascribed to the onset of CAF order [17]. Likewise, kinks were found at 17 K, 21 K and 25 K for $x = 0.33, 0.5$ and $0.67$, respectively. The linear decrease in the kink temperature (see Fig. 3(b)) strongly suggests that those kinks are due to magnetic order as well. In figure 4, the magnetization curves for $x \geq 0.33$ show a linear increase over a wide field range. Although the magnetizations do not saturate up to 60 T applied in this study, one can say that the saturation field would become smaller with decreasing $x$, implying a weaker superexchange constant for Cu-Cl-Cu than for Cu-Br-Cu. On the other hand, the magnetization curve for $x = 0.05$ is considerably different from those for $x \geq 0.33$.

As shown Figs. 3 (a) and 4, both $\chi(T)$ and $M(H)$ for $x = 0.05$, behave quite similarly to those for $x = 0$ [14, 15]; $\chi(T)$ is characterized by a broad maximum at about 15 K and a substantial decrease in $\chi(T)$ below this temperature, and the concave behavior in $M(H)$ around 10 T (see the inset of Fig. 4) is reminiscent of a transition from the spin-singlet state to the field-induced antiferromagnetic state. However, recent $\mu$SR experiments [21] discovered the presence of magnetic order of some kind in this material. Accordingly, in order to obtain insight into the ground state of $x = 0.05$, we performed neutron powder diffraction measurements. Figure 5(a) shows the intensity difference, $I$ (3.5 K) − $I$ (18 K), in the 2theta range from 16° to 20°. We observed a relatively weak magnetic peak corresponding to the (1/2 0 1/2) reflection and the observed peak width was resolution limits. Hence, it is natural to assume that this material has the same CAF order as (CuBr)LaNb$_2$O$_7$ ($x$ = 1) [17]. From the temperature dependence of this magnetic reflection (Fig. 5(b)), we estimated $T_N$ as 7 K, which again agrees with the µSR result. From the data at 3.5 K, the magnitude of the magnetic moment in the ordered state was roughly estimated as 0.2(1) $\mu_B$, where the crystal structure of (CuCl)LaNb$_2$O$_7$ determined by X-ray diffraction at room temperature [13] and the magnetic structure with the moments aligned parallel to the $b$ axis as found in (CuBr)LaNb$_2$O$_7$ [17] were assumed. The estimated ordered moment is significantly smaller than 0.6 $\mu_B$ for $x$ = 1. It is also evidenced from the µSR measurements [21] that the magnetic moment for $x$ = 0.05 is considerably smaller than that for $0.33 \leq x \leq 1$. The observations of long-range magnetic order induced by substituting only 5% Br, and the significantly reduced magnetic moment for $x$ = 0.05 indicate that (CuCl)LaNb$_2$O$_7$ is located in the vicinity of the quantum phase boundary to the ordered state.

Contrasting behavior has been observed in the Nb-site substituted system, (CuCl)La(Nb$_{1-y}$Ta$_y$)$_2$O$_7$ [21, 24], in which the spin-singlet state is robust against cationic substitution. Substantial substitution on the order $y \sim 0.4$ is needed to induce CAF magnetic order, but with coexistence of the ordered and disordered phases extending up to $y$ = 1. It is notable that the effect of chemical disorder in the magnetic layer Cu$X$ is larger than that in the non-magnetic layer La$B^{5+}_2$O$_7$ ($B^{5+}$: Nb$^{5+}$ or Ta$^{5+}$ ($d^0$)) So far, there has been much theoretical work on the rapid breaking of spin-singlet dimer upon the addition of vacancies in a variety of spin-gapped models, a procedure which is considered relevant to creating free spins. Some theories showed that the staggered spin-spin correlations are enhanced in the vicinity of spin vacancies [25−29], which accounts for the emergence of the ordered phase in cationic-substituted 1D magnetic materials CuGeO$_3$ and SrCu$_2$O$_3$ [9, 10]. In the 2D orthogonal dimer SrCu$_2$(BO$_3$)$_2$, staggered magnetization develops around the triplet excitation in the 1/8 plateau phase [5]. Thus, in spin-singlet compounds, independent of their dimensionality, any sort of local perturbations such as defects or chemical randomness can potentially induce staggered magnetizations, leading to a magnetically ordered state. By analogy, the creation of the magnetic order by a tiny amount of Br substitution in (CuCl)LaNb$_2$O$_7$ may be associated with staggered correlations induced around Br ions. A rapid depolarization of µSR signal observed in the 5% Br system [21] may be related to spatial variation of staggered ordered moment size. Earlier µSR spectra in CuGeO$_3$ doped with Zn and Si exhibit fast relaxation due to this effect [9]. Further experiments are necessary to clarify the local state.

In conclusion, we succeeded in synthesizing the solid solution system (CuCl$_{1-x}$Br$_x$)LaNb$_2$O$_7$ ($0 \leq x \leq 1$) and investigated its magnetic properties by means of susceptibility, magnetization and powder neutron diffraction. These experimental results reveal that switching of the spin-singlet state in (CuCl)LaNb$_2$O$_7$ to a collinear-type magnetic ordered state occurs by as little as 5 % substitution of Br, which is in stark contrast to the Ta-for-Nb-substitution systems. It is likely that the chemical disorder in the magnetic layer Cu$X$ has more influence on the magnetic properties than that in the non-magnetic layer La$B_2$O$_7$.

The chemistry of present material has ability to yield a big family represented as ($MX$)$A_{n-1}B_n$O$_{3n+1}$ ($M$: divalent transition-metal, $A$: alkali, alkali-earth, and rare-earth metals, $n$: integer) [23], and those materials provide a variety of magnetic states, depending on tunable parameters. For example, it was shown that substituting $A$-site in (CuBr)$A_2$Nb$_3$O$_{10}$ controls the width of the 1/3 magnetization plateau [30]. Here we note that ion-exchange reactions to yield ($MX$)$A_{n-1}B_n$O$_{3n+1}$ rely on the stability of the byproduct alkali halides, in the present case RbCl and RbBr, relative to transition-metal halides, so anionic substitution should be generally applied to the entire ($MX$)$A_{n-1}B_n$O$_{3n+1}$ systems. It would be interesting to see, e.g., how anionic substitution for (CuBr)Sr$_2$Nb$_3$O$_{10}$ influences the 1/3 plateau phase.

This work was supported by Grants-in-Aid for Science Research on Priority Areas "Novel States of Matter Induced by Frustration" (No. 19052004) from the Ministry of Education, Culture, Sports, Science and Technology of Japan, the Japan-U.S. Cooperative Science Program from JSPS of Japan and NSF (No. 14508500001), and by the Global COE program International Center for Integrated Research and Advanced Education in Material Science, Kyoto University, Japan. The work at Columbia University was supported by US NSF grants DMR-05-02706 and DMR-08-06846. One of the authors (Y. T.) was supported by the Japan Society for the Promotion of Science for Young Scientists.

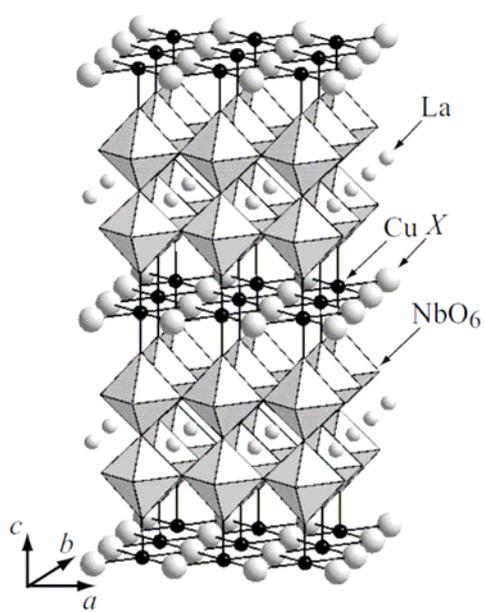

**Figure 1.**

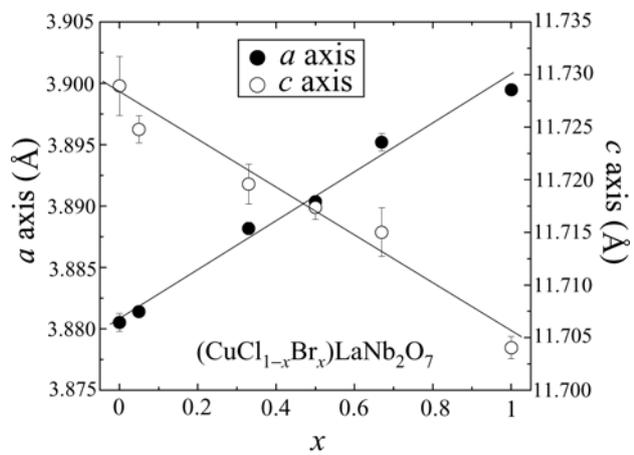

**Figure 2.**

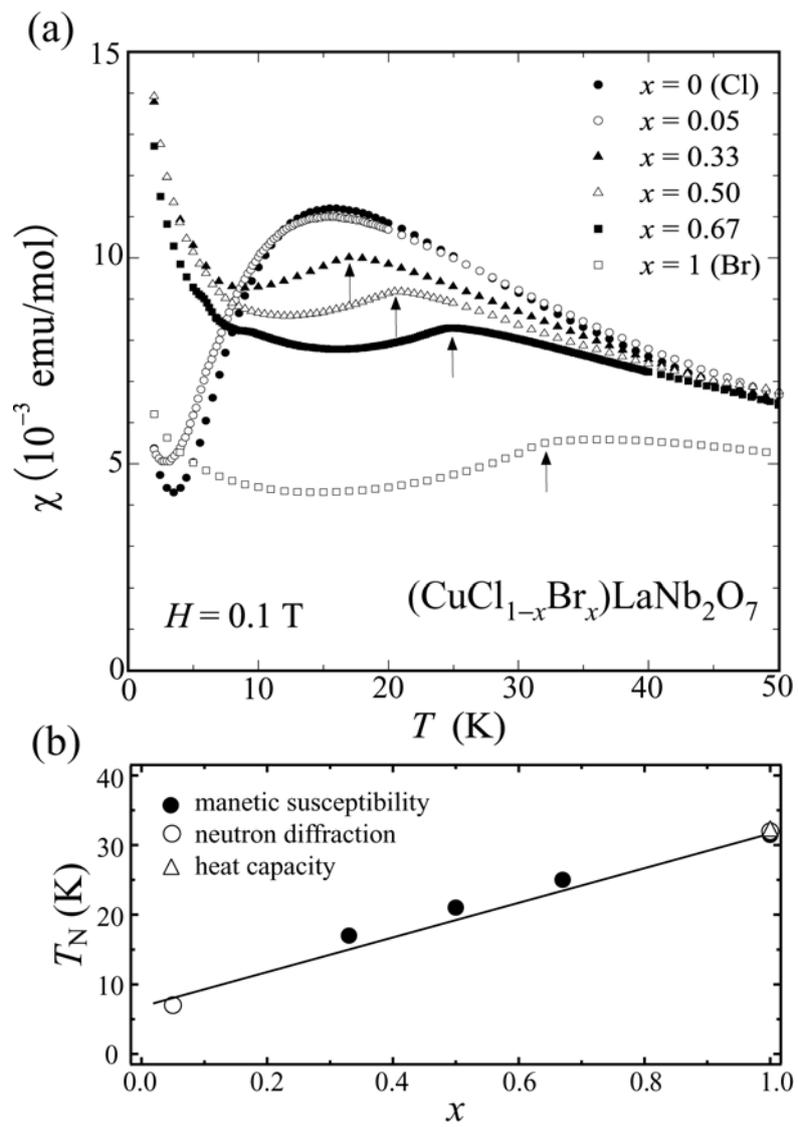

**Figure 3.**

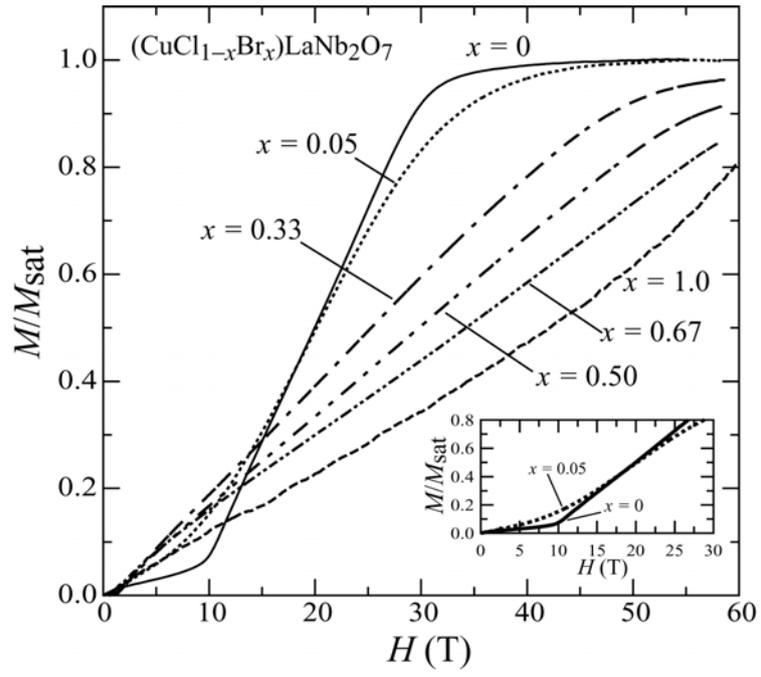

**Figure 4.**

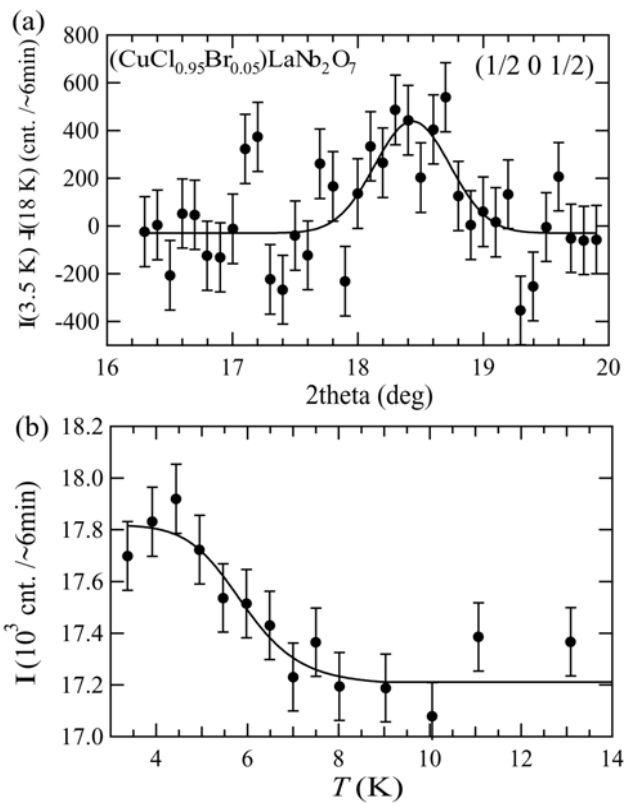

**Figure 5.**

Figure 1. Crystal structure of (Cu$X$)LaNb$_2$O$_7$ ($X$ = Cl, Br).

Figure 2. Crystallographic unit-cell parameters as a function of Br concentration ($x$) in (CuCl$_{1-x}$Br$_x$)LaNb$_2$O$_7$.

Figure 3. (a) Magnetic susceptibilities for (CuCl$_{1-x}$Br$_x$)LaNb$_2$O$_7$ measured at $H$ = 0.1 T. Arrows point the kink temperatures. (b) The $x$ dependence of $T_N(x)$, determined by magnetic susceptibility (solid circles), neutron diffraction (open circles) and heat capacity (triangle) [16].

Figure 4. Magnetization curves for (CuCl$_{1-x}$Br$_x$)LaNb$_2$O$_7$ measured at 1.3 K. The inset shows the magnetizations for $x$ = 0 and 0.05 below $H$ = 30 T.

Figure 5. (a) The difference plot of neutron scattering intensity for the $x$ = 0.05 sample, $I$(3.5 K) – $I$(18 K), corresponding to the (1/2 0 1/2) magnetic reflection. The solid curve is the single-component Gaussian fit. (b) The temperature dependence of the (1/2 0 1/2) intensity. The solid line is a guide to the eyes.